# Agent-based Simulation of Human Movement Shaped by the Underlying Street Structure

Bin Jiang[(1)] and Tao Jia[(2)]

[(1)] Department of Technology and Built Environment, Division of Geomatics
University of Gävle, 801 76 Gävle, Sweden
Email: bin.jiang@hig.se,

[(2)]Future Position X, Box 975, 801 33 Gävle, Sweden
Email: jiatao83@163.com

*(September 2009)*

**Abstract**
Relying on random and purposive moving agents, we simulated human movement in large street networks. We found that aggregate flow, assigned to individual streets, is mainly shaped by the underlying street structure, and that human moving behavior (either random or purposive) has little effect on the aggregate flow. This finding implies that given a street network, the movement patterns generated by purposive walkers (mostly human beings) and by random walkers are the same. Based on the simulation and correlation analysis, we further found that the closeness centrality is not a good indicator for human movement, in contrast to a long standing view held by space syntax researchers. Instead we suggest that Google's PageRank, and its modified version - weighted PageRank, betweenness and degree centralities are all better indicators for predicting aggregate flow.

## 1. Introduction

How do people move around in street networks? Why do some streets tend to be more attractive than other streets? Does human traveling behavior have an effect on human flow distribution? These are some fundamental issues that face the practitioners and researchers in urban planning and design as well as in traffic and transport management. In the space syntax literature (e.g., Hillier and Hanson 1984, Hillier 1996, Hillier et al. 1993, Penn et al. 1998, Turner 2007b), it is found that human movement is predictable for individual streets. To put it simply, well connected streets tend to attract more traffic than less connected streets. This sounds intuitive enough. However, what is counter-intuitive is that such a simple model, purely from a topological perspective without considering geometric distance and other factors such as building height, land use and population density, can predict aggregate flow or human movement to a fairly high percentage. At a conservative level, about 70% of traffic flow can be predicted by just looking at the underlying street structure.

What do we mean by street structure then? It refers to a street-street relationship as to what street connects to what other streets. More specifically, the street-street relationship is represented by a connectivity graph, consisting of nodes representing individual streets, and links if the two streets are connected to each other (*c.f.*, Figure 1 for illustration). The street structure is then defined and characterized by a range of graph theoretic metrics (*c.f.* Appendix) for ranking the importance of individual streets. For instance, a universal street structure states that 80% of streets have a connectivity (Note: connectivity and degree centrality are interchangeably used in the text) of less than four (an average connectivity of streets), while 20% of streets have a connectivity greater than the average (Jiang 2007). Following the street structure, over 80% of traffic occurred in the 20% of well connected streets; while only 20% of traffic occurred in the 80% less connected streets (Jiang 2009). This is a movement pattern that is shaped by the universal street structure.

Now that human movement is mainly shaped by the underlying street structure, human traveling behavior would have little effect on movement patterns. This kind of human movement is termed natural movement (Hillier et al. 1993). However, while elaborating on why space syntax works, in particular why natural movement can be captured by local integration, Hillier and his colleagues (Hillier 1999, Hillier and Iida 2005) attempted to seek an answer from the human side, such as people's conceptualization of distance, and people's tendency to minimize trip length or maximize trip efficiency. This is implausible and rather odd reasoning. Instead, we will argue that it is not the people but the underlying street structure that determines the movement patterns. Even if we put some random walkers in a street network, the ultimate movement pattern formed by the random walkers would be the same as that by human beings.

A recent study by Jiang, Yin and Zhao (2009) has made an effort toward illustrating the fact that goal-directed human movement has little effect on traffic flow distribution. Using massive GPS data captured from 50 taxicabs and related customer data, the study made a careful observation about real-world human movement approximated by the movement of taxicabs. Interestingly, random walkers can replicate the observed human movement patterns. However, the movement patterns captured by the taxicabs are substantially biased, since most taxicabs go to or come out from a very few hotspots such as the central station, the city center, and the hospital. This is the very reason why we imposed two additional constraints while simulating random walkers' movement: (1) the random walkers start from one of the origins or destinations, (2) the random walkers' traveling time is power law distributed. The simulated random movement can reproduce the observed movement patterns. In this paper, we are going to release these two constraints. That is, the random walkers can start from any location (not necessarily one of the origins or destinations), and there is also no need for the traveling time to be power law distributed. Eventually, the simulated aggregate random movement is still highly correlated to the street structure.

This paper is further motivated by challenging the conventional wisdom in the literature that closeness centrality (or global integration in terms of space syntax, so both are interchangeably used in the paper) is a default indicator for aggregate flow. We will provide statistical evidence to support our argument that closeness centrality is not a good indicator. This is also the conjecture developed from the previous study by Jiang (2009), where weighted PageRank scores were found to better correlate to traffic flow than local integration.

The remainder of this paper is structured as follows. Section 2 introduces the concept of streets, two kinds of moving agents or walkers and two different ways of counting aggregate flow assigned to individual streets. Section 3 describes in detail the data sources and simulation environments. In section 4, we examine the experimental results to support our arguments: (1) human moving behavior has little effect on the aggregate flow, and thus the movement patterns of human beings and the random walkers, given a same street network, are essentially the same, (2) closeness centrality is not a good indicator for aggregate flow, and therefore we suggest some alternatives. Finally section 5 concludes the paper and points to future work.

**2. Streets, moving agents and aggregate flow**
The movement we deal with in this paper is constrained by streets or street networks, and it is in contrast to free movement as for instance in a public square or inside a building complex (e.g., Turner 2007). Streets are defined as semantically meaningful units, distinct from street segments between two adjacent junctions. Under this street definition, streets are represented in various forms: represented by individual axial lines – the longest visibility lines cutting across open space between buildings or street blocks, identified by unique names – named streets (Jiang and Claramunt 2004), and by self-organized processes – natural streets (Jiang and Liu 2009, Jiang, Zhao and Yin 2008). It is worth noting that the natural streets are generated based on the Gestalt principle of good continuity, and they are also referred to as strokes (Thomson 2003), or continuity lines (Figueiredo and Amorim 2005). Interestingly, it is found based on massive GPS data that human routes seem to follow the same principle of good continuity as well (Turner 2009). No matter what street forms are adopted, they all form an interconnected whole, represented by a connectivity graph, consisting of nodes representing individual streets and links if the two streets are connected to each other (Figure 1).

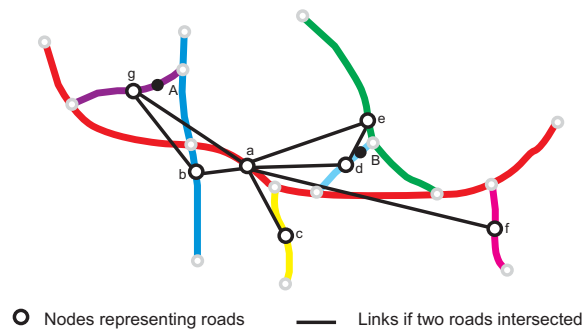

Figure 1: (Color online) A connectivity graph based on street-street interaction (remove node numbers)
(Note: Colors are used to indicate different streets or how different segments are merged together to form individual streets. This color usage is different from Figure 2, where similar colors indicate some metric scores)

Two kinds of moving agents are implemented to simulate movement in street networks (*c.f.* Appendix B for algorithms). Both are inspired by PageRank's random surfer model, different in essence from the agents used in Penn and Dalton (1994), which move randomly from junction to junction or follow shortest path routes. The first is random walkers that can hop arbitrarily from one street to another. The hopping behaviour is defined at a topological level with the connectivity graph. On the other hand, the simulation has yet to be based on a geometric level to mimic the network constrained movement. That is, within an individual street, the random walkers persistently move along the street until they reach the next street that intersects with the current one. As soon as the random walkers reach the intersection, a decision as to which will be the next street has to be made (again randomly decided, but sometimes with a higher priority to highly connected streets), and then move toward the next intersection, and so on and so forth. The random behaviour is obviously occurring at the topological rather than the geometric level.

The second type of moving agents is more realistic and goal oriented – purposive walkers. Suppose that a person wants to go from location A to B (Figure 1), located respectively in street g and street d, he/she would walk (1) along street g and toward the streets intersection of g and a, (2) along street a, and toward the streets intersection of a and d, and (3) along street d, and toward location B. This is just for illustration purposes. In a large street network, numerous purposive walkers would randomly decide their goals and then target them purposively. Once they reach their first goals, they randomly choose their second goals and target them purposively and so on and so forth. The purposive walkers have no preference in choosing destinations or goals. This is the first type of purposive walkers. The second type of purposive walkers give a high priority (with a probability of 80%) to closer locations within two steps of topological distance, while a low priority (with a probability of 20%) to farther locations beyond two steps. This is based on the observation that most people travel only short distances, while a few regularly move over hundreds of miles (Gonzalez, Hidalgo and Barabási 2008). We used topological distance rather than geometric distance to characterize the second type of purposive walkers. This is due to our belief that topology matters more than geometry in shaping movement patterns.

How should one count aggregate flow assigned to individual streets? A first approach is called gate counts, i.e., set a gate count in each street segment and count traffic flow passing through it. Space syntax relies on this method for counting traffic flow, and all the gates' flow along one street (or axial line) are aggregated to be that of the street. This way of counting is fairly accurate if every street segment has a gate. Note that the gate counts approach cannot differentiate traffic between two street segments of different lengths, as it counts only once while a vehicle or a person passes through a gate. However, real traffic flow should refer to the extent that vehicles (or persons equivalently) occupy the streets. In this sense, given that a vehicle passes two street segments of different lengths, the longer street segment would have more traffic than the shorter one. This is because more footprints are left in the longer street segment than in the shorter one.

A second approach to counting aggregate flow is called footprint counts. The basic idea of footprint counts is to count how many footprints are left with movement trajectories. Footprint counts are hard to implement in reality, unless all vehicles running in the street network for example have a GPS receiver. For the vehicles with varying speeds (which is the case in reality), the number of footprints must be adjusted to remove the speed effect, i.e., the faster the speed, the fewer footprints are left. For vehicles with a consistent speed (which is the likely case in a simulation), the number of footprints reflect precise traffic flow. This way of counting can easily be done in simulated environments. For the footprint counts, there are two different ways of counting flow: one at the topological level and another at the geometric level. People coming to one street (rather than street segment) are counted once, no matter how long they walk along the street. This is thus counted at the topological level. On the other hand, the number of footprints recorded at every time interval (e.g., 2 seconds) reflects the aggregate flow at the geometric level. The topological flow is mainly used to determine that a simulation reaches to a saturated status (i.e., when the topological flow is well correlated to the computed PageRank scores), while the geometric flow reflects the simulated flow.

**3. Data sources and simulation environments**
We adopted two data sources respectively representing two forms of streets: axial lines and natural streets. The data sources are publicly available. The first is the London data, including both axial lines and observed traffic flow. The data have been used in previous studies (e.g., Hillier et al. 1993, Jiang 2009), and are downloadable at http://eprints.ucl.ac.uk/1398/. Instead of using the whole of London (~300 km$^2$), three local areas in the centre are sampled for detailed observation studies. The three areas are Barnsbury, Clerkenwell, and South Kensington with 1915, 3622, and 2742 axial lines respectively (Figure 2a). Traffic data of both pedestrians and vehicles are observed in the four sites within the three areas: two for the first two areas Barnsbury and Clerkenwell, and another two for the third area South Kensington. Although the traffic data have been used in previous studies, we have some doubts on how the observed data reflect a true picture of traffic in reality. Our major concern is due to the allocation of the gates. There are some missing gates for some axial lines, implying a possible bias.

The second data were previously used in the study by Jiang, Yin and Zhao (2009), and publicly available at http://fromto.hig.se/~bjg/PREData/. We mainly count on the street network rather than the traffic data of the 50 taxicabs, since the traffic data do not represent a true picture of traffic in reality as elaborated earlier in the paper. This is a very large street network (~1000 km$^2$), including 4056 natural streets generated from 10 439 street segments across the four cities or towns in the middle of Sweden: Gävle, Sandviken, Storvik, and Hofors. For the sake of convenience, we call the data source Gävle data.

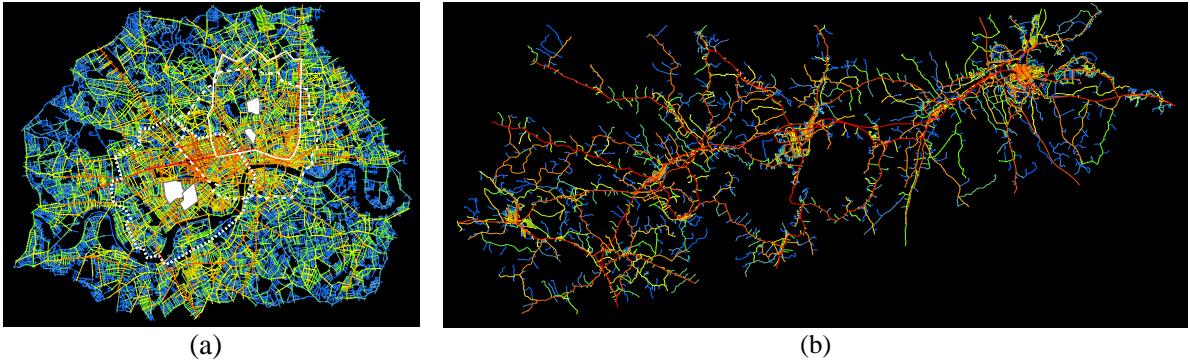

(a)                  (b)

Figure 2: (Color online) Data sources for the agent-based simulation: (a) London axial map, three sampled areas – Barnsbury (solid), Clerkenwell (dashed), South Kensington (dotted), and four observed sites (patches highlighted), and (b) street map of the four cities in the middle of Sweden (Note: the streets are visualized by spectral color ranging from red (highest) to blue (lowest) according to weighted PagaRank scores to be introduced in the following section)

Built on NetLogo (Wilensky 1999) and ArcGIS, both random and purposive moving agents are implemented to interact with the underlying streets to generate movement. NetLogo is a freeware mainly developed for agent-based simulation, with a very good visualization capability and easy-to-use interface; while ArcGIS is a commercial GIS product that can accommodate very large street networks. In many aspects, the two software platforms complement each other. In this study, we put the London data in NetLogo and the Gävle data in ArcGIS with our simulation programs developed to keep control of the moving behaviour and compute related parameters. It should be noted that the simulation is very time consuming even with state-of-the-art personal computers. This is particularly true for the purposive agents, which have to compute the shortest paths. For the London cases, we set up 200 agents for each type of the three walkers, running in the networks for 50k ticks; while for the Gävle case, we set up 500 agents for each type, running in the network for 100k ticks. Furthermore, for every 1k ticks the simulation programs have to compute an R square value for metric-flow correlations.

**4. Experimental results**

To characterize street structure, we adopt seven metrics including weighted PageRank (WPR), PageRank (PR), connectivity (Cnt), control (Ctr), betweenness (Btw), local integration (Ltg), and global integration (Gtg) (*c.f.* Appendix for an introduction). We examine the metric-flow correlations, in order to (1) see how aggregate flow can be captured by the metrics, or equivalently, to see how movement patterns are shaped by the underlying street structure, and (2) check which metrics are the best indicators for predicting traffic flow at a collective level.

Using the London data, in particular the observed traffic flow of both pedestrians and vehicles, we correlate the aggregate flow of the individual axial lines to the seven metrics. It seems that all seven metrics have a good correlation with the aggregate flow (Table 1). On the other hand, both Ltg and WPR are outperformed at the same level (60% for pedestrians and 70% for vehicles), while Ctr and Gtg seem to be the worst in terms of metric-flow correlation. All the metrics, except for Ltg and Gtg, exhibit a power law distribution (Jiang, Zhao and Yin 2008), so before computing the R square values shown in Table 1, we took quartic root for all the metrics whose value distributions are skewed. This implies that the correlation is based on two variables that are approximately normally distributed.

Table 1: R square values between the metrics and the observed traffic flow of both pedestrians and vehicles

|     | Barnsbury | | Clerkenwell | | S.Kensington | | Knightsbridge | | Mean | |
|-----|-----|-----|-----|-----|-----|-----|-----|-----|-----|-----|
|     | Ped | Veh | Ped | Veh | Ped | Veh | Ped | Veh | Ped | Veh |
| WPR | 0.78 | 0.64 | 0.58 | 0.69 | 0.64 | 0.80 | 0.54 | 0.65 | **0.64** | **0.69** |
| PR  | 0.75 | 0.60 | 0.52 | 0.64 | 0.65 | 0.79 | 0.48 | 0.58 | 0.60 | 0.65 |
| Cnt | 0.75 | 0.60 | 0.52 | 0.64 | 0.65 | 0.79 | 0.48 | 0.58 | 0.60 | 0.65 |
| Ctr | 0.65 | 0.49 | 0.40 | 0.43 | 0.49 | 0.66 | 0.36 | 0.42 | 0.47 | 0.50 |
| Btw | 0.71 | 0.61 | 0.36 | 0.71 | 0.55 | 0.67 | 0.55 | 0.63 | 0.54 | 0.65 |
| Ltg | 0.78 | 0.64 | 0.58 | 0.72 | 0.66 | 0.76 | 0.50 | 0.66 | **0.63** | **0.70** |
| Gtg | 0.60 | 0.56 | 0.39 | 0.69 | 0.42 | 0.62 | 0.51 | 0.50 | 0.48 | 0.59 |

The metric-flow correlation coefficients in Table 1 provide a benchmark for simulated flow that is captured by the original gate counts for both pedestrians and vehicles. It is in this sense that the simulated walkers act as either pedestrians or vehicles. In fact, we did not differentiate pedestrians and vehicles in our simulations. For comparison purposes, we use the same gates (as in reality for observing flow of pedestrians and vehicles) to capture simulated flow and then aggregate to individual axial lines. It is indeed true that the simulated aggregate flow can be captured by the metrics as well. Taking Barnsbury for example, the R square value reaches a stable point when simulation time approaches 50k ticks (Figure 3). Overall, the correlation coefficient is over 60%, while the highest coefficient can reach 90%. Comparing pedestrians (Figure 3 a, b, and c) and vehicles (Figure 3, d, e, and f), it appears that all the metrics capture pedestrian flow a bit better than vehicle flow. But the difference is small.

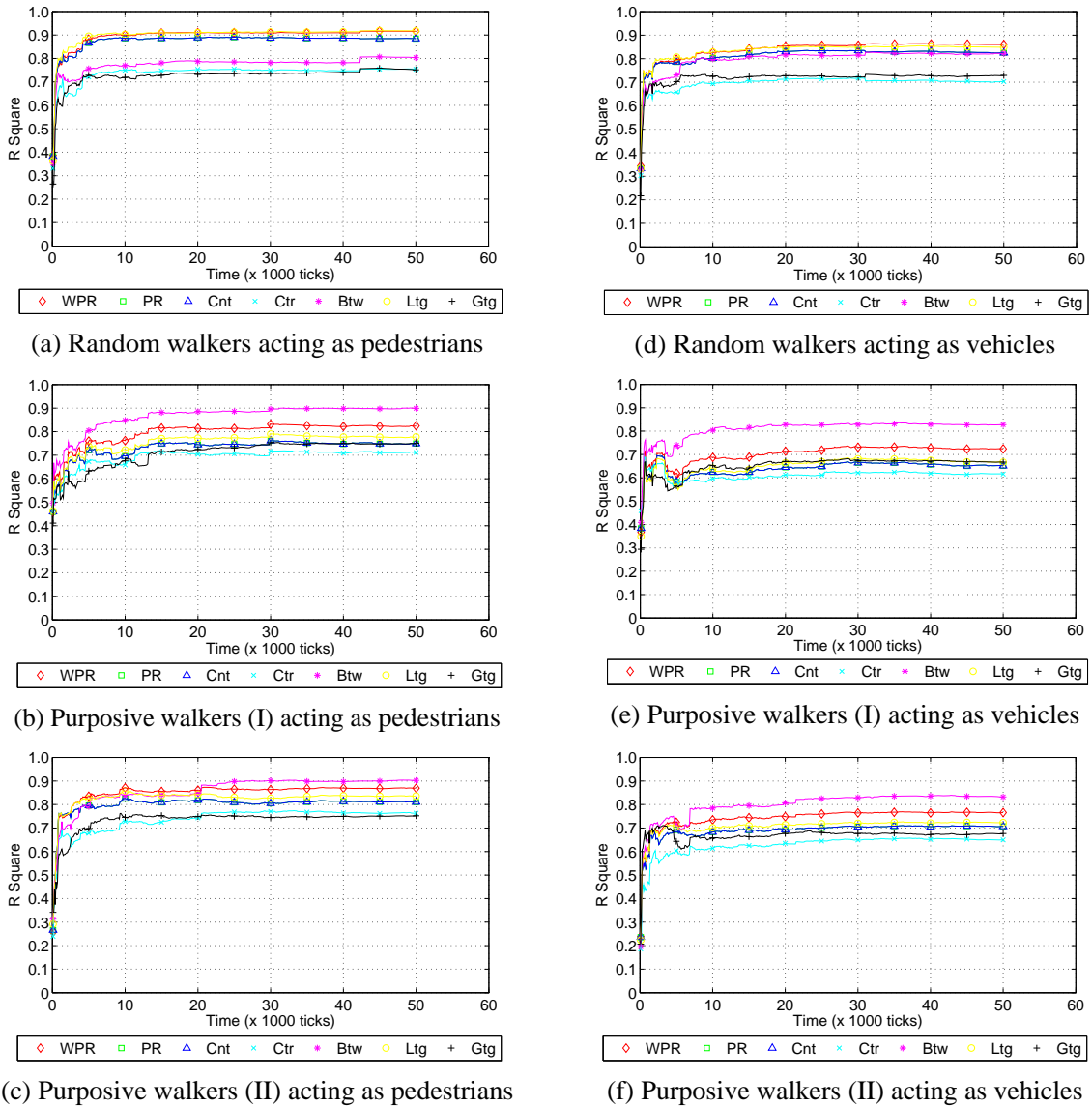

(a) Random walkers acting as pedestrians    (d) Random walkers acting as vehicles

(b) Purposive walkers (I) acting as pedestrians    (e) Purposive walkers (I) acting as vehicles

(c) Purposive walkers (II) acting as pedestrians    (f) Purposive walkers (II) acting as vehicles

Figure 3: (Color online) Simulated aggregate flow captured by the original gate counts is highly correlated to the seven metrics based on the observed site within Barnsbury with moving agents acting as pedestrians (a), (b) and (c), and as vehicles (d), (e) and (f)

Let us consider the R square values at the time instant of 50k ticks for four observed sites together (Table 2). We again see that most of metrics are highly correlated to the simulated aggregate flow, nearly all over 60%, and some approaching 90%. This indeed illustrates the fact that human moving behaviour has little effect on the aggregate flow. Taking a closer look, we will find some differences, in particular between the two types of purposive walkers, i.e., the second type of purposive walkers is slightly outperformed by the first type of purposive walkers. However, the difference is not so substantial in terms of the absolute values and in comparison with the case of random walkers.

Table 2: Metrics-flow correlation (R square values) for simulated flow in the four observed sites using the original gates (Note: Unlike the observed flow, simulated flow shows no difference between pedestrians and vehicles; the ped and veh in this table indicate gate counters initially set for pedestrians and vehicles in reality)

|  |  | Barnsbury | | Clerkenwell | | S.Kensington | | Knightsbridge | | Mean | |
|---|---|---|---|---|---|---|---|---|---|---|---|
|  |  | Ped | Veh | Ped | Veh | Ped | Veh | Ped | Veh | Ped | Veh |
| Random walkers | WPR | 0.92 | 0.86 | 0.90 | 0.90 | 0.91 | 0.91 | 0.87 | 0.89 | **0.90** | **0.89** |
|  | PR | 0.88 | 0.82 | 0.88 | 0.89 | 0.90 | 0.92 | 0.83 | 0.83 | 0.87 | 0.87 |
|  | Cnt | 0.88 | 0.82 | 0.88 | 0.89 | 0.90 | 0.92 | 0.83 | 0.83 | 0.87 | 0.87 |
|  | Ctr | 0.75 | 0.70 | 0.61 | 0.64 | 0.67 | 0.71 | 0.61 | 0.64 | 0.66 | 0.67 |
|  | Btw | 0.80 | 0.82 | 0.63 | 0.70 | 0.72 | 0.74 | 0.76 | 0.83 | 0.73 | 0.77 |
|  | Ltg | 0.92 | 0.85 | 0.91 | 0.91 | 0.87 | 0.90 | 0.85 | 0.85 | **0.89** | **0.88** |
|  | Gtg | 0.75 | 0.73 | 0.57 | 0.63 | 0.60 | 0.67 | 0.33 | 0.39 | 0.56 | 0.60 |
| Purposive walkers (I) | WPR | 0.82 | 0.72 | 0.63 | 0.60 | 0.81 | 0.81 | 0.65 | 0.64 | **0.79** | **0.70** |
|  | PR | 0.75 | 0.65 | 0.62 | 0.59 | 0.78 | 0.77 | 0.57 | 0.54 | **0.68** | **0.64** |
|  | Cnt | 0.75 | 0.65 | 0.62 | 0.59 | 0.78 | 0.77 | 0.57 | 0.54 | **0.68** | **0.64** |
|  | Ctr | 0.71 | 0.62 | 0.52 | 0.51 | 0.74 | 0.74 | 0.47 | 0.43 | 0.61 | 0.58 |
|  | Btw | 0.90 | 0.83 | 0.87 | 0.85 | 0.88 | 0.88 | 0.83 | 0.84 | **0.87** | **0.85** |
|  | Ltg | 0.78 | 0.67 | 0.62 | 0.57 | 0.70 | 0.72 | 0.56 | 0.54 | 0.67 | 0.63 |
|  | Gtg | 0.75 | 0.67 | 0.51 | 0.51 | 0.67 | 0.71 | 0.44 | 0.48 | 0.59 | 0.59 |
| Purposive walkers (II) | WPR | 0.87 | 0.77 | 0.75 | 0.68 | 0.85 | 0.85 | 0.69 | 0.72 | **0.79** | **0.75** |
|  | PR | 0.81 | 0.71 | 0.70 | 0.66 | 0.83 | 0.82 | 0.63 | 0.63 | **0.74** | **0.71** |
|  | Cnt | 0.81 | 0.71 | 0.70 | 0.66 | 0.83 | 0.82 | 0.63 | 0.63 | **0.74** | **0.71** |
|  | Ctr | 0.76 | 0.65 | 0.54 | 0.55 | 0.75 | 0.76 | 0.52 | 0.50 | 0.64 | 0.61 |
|  | Btw | 0.90 | 0.83 | 0.82 | 0.85 | 0.87 | 0.88 | 0.84 | 0.87 | **0.86** | **0.86** |
|  | Ltg | 0.84 | 0.72 | 0.71 | 0.65 | 0.76 | 0.77 | 0.62 | 0.64 | 0.73 | 0.70 |
|  | Gtg | 0.75 | 0.68 | 0.52 | 0.55 | 0.68 | 0.72 | 0.41 | 0.43 | 0.59 | 0.59 |

So far the simulated aggregate flow is captured by the gate counts originally set for the observation. There are many missing gates, which should be added for a better observation. After adding the missing gates, we then run the simulations, and found that metric-flow correlations are substantially improved (Table 3). Surprisingly, most of the R square values reach over 90% for random walkers and over 80% for purposive walkers. From that, we can estimate the extent to which the missing gate counts have a biased effect on the metric-flow correlations.

Through the above examination, we have seen that local integration is one of the best metrics, but far from the best metric in capturing the simulated aggregate flow as the literature indicated. For random walkers, WPR is almost as good as Ltg; while for purposive walkers either (I) or (II), there are many metrics that are better correlated to traffic than Ltg (see those figures highlighted in bold with Table 2 and 3). It is striking enough that Gtg performs very badly in terms of metric-flow correction. Therefore Gtg is NOT a good indicator at all for predicting traffic flow. What about Ltg? It looks like a good indicator, but far from the best one from the four observed sites. A similar observation remains valid for the three London local areas as well as for the entire Gävle area (Table 4 using footprint counts, and Table 5 using gate counts). However, it is striking that with the Gävle network (see column Gävle), WPR outperforms all types of moving agents. This is different from the London counterparts (see column Mean), where Btw appears the best for purposive walkers, and WPR is the best for random walkers. Furthermore, Ltg in the Gävle case performs pretty poorly in comparison with the London cases. Note that for the simulations with the Gävle data, it took a much longer time to saturate the simulation, i.e., 100k ticks with 500 agents for each one of the three different types of walkers. As for the reason why there is such a great difference between the London cases and the Gävle case, we are not sure yet. What seems to be consistent is that Gtg is the poorest indicator and that Ltg is far from the best indicator.

Table 3: Metrics-flow correlations (R square value) for simulated flow in the four observed sites using the original and missing gates together

|  |  | Barnsbury | | Clerkenwell | | S.Kensington | | Knightsbridge | | Mean | |
|---|---|---|---|---|---|---|---|---|---|---|---|
|  |  | Ped | Veh | Ped | Veh | Ped | Veh | Ped | Veh | Ped | Veh |
| Random walkers | WPR | 0.98 | 0.98 | 0.97 | 0.98 | 0.97 | 0.97 | 0.96 | 0.97 | **0.97** | **0.97** |
|  | PR | 0.96 | 0.98 | 0.95 | 0.96 | 0.96 | 0.97 | 0.94 | 0.95 | **0.95** | **0.97** |
|  | Cnt | 0.96 | 0.98 | 0.95 | 0.96 | 0.96 | 0.97 | 0.94 | 0.95 | **0.95** | **0.97** |
|  | Ctr | 0.83 | 0.87 | 0.70 | 0.73 | 0.76 | 0.82 | 0.77 | 0.78 | 0.77 | 0.80 |
|  | Btw | 0.85 | 0.87 | 0.68 | 0.74 | 0.84 | 0.81 | 0.78 | 0.78 | 0.79 | 0.80 |
|  | Ltg | 0.97 | 0.97 | 0.96 | 0.96 | 0.91 | 0.91 | 0.90 | 0.92 | **0.94** | **0.94** |
|  | Gtg | 0.78 | 0.78 | 0.60 | 0.66 | 0.68 | 0.68 | 0.31 | 0.29 | 0.59 | 0.60 |

|  |  | | | | | | | | | |
|---|---|---|---|---|---|---|---|---|---|---|
| Purposive walkers (I) | WPR | 0.88 | 0.86 | 0.81 | 0.79 | 0.89 | 0.88 | 0.79 | 0.83 | **0.84** | **0.84** |
|  | PR  | 0.82 | 0.82 | 0.76 | 0.78 | 0.86 | 0.85 | 0.74 | 0.75 | **0.80** | **0.80** |
|  | Cnt | 0.82 | 0.82 | 0.76 | 0.78 | 0.86 | 0.85 | 0.74 | 0.75 | **0.80** | **0.80** |
|  | Ctr | 0.77 | 0.81 | 0.62 | 0.67 | 0.79 | 0.81 | 0.65 | 0.64 | 0.71 | 0.73 |
|  | Btw | 0.92 | 0.93 | 0.89 | 0.93 | 0.92 | 0.94 | 0.90 | 0.92 | **0.91** | **0.93** |
|  | Ltg | 0.83 | 0.80 | 0.76 | 0.76 | 0.77 | 0.78 | 0.68 | 0.72 | **0.76** | **0.77** |
|  | Gtg | 0.79 | 0.74 | 0.60 | 0.61 | 0.77 | 0.75 | 0.41 | 0.45 | 0.64 | 0.64 |
| Purposive walkers (II) | WPR | 0.93 | 0.91 | 0.87 | 0.85 | 0.92 | 0.91 | 0.85 | 0.88 | **0.89** | **0.89** |
|  | PR  | 0.88 | 0.88 | 0.83 | 0.83 | 0.90 | 0.90 | 0.83 | 0.83 | **0.86** | **0.86** |
|  | Cnt | 0.88 | 0.88 | 0.83 | 0.83 | 0.90 | 0.90 | 0.83 | 0.83 | **0.86** | **0.86** |
|  | Ctr | 0.82 | 0.84 | 0.66 | 0.69 | 0.81 | 0.85 | 0.71 | 0.71 | 0.75 | 0.77 |
|  | Btw | 0.92 | 0.93 | 0.86 | 0.89 | 0.92 | 0.93 | 0.88 | 0.91 | **0.89** | **0.91** |
|  | Ltg | 0.88 | 0.86 | 0.83 | 0.81 | 0.82 | 0.83 | 0.76 | 0.78 | **0.82** | **0.82** |
|  | Gtg | 0.78 | 0.76 | 0.61 | 0.63 | 0.70 | 0.76 | 0.38 | 0.39 | 0.62 | 0.64 |

Table 4: R square values between the metrics and the simulated flow (using footprint counts) for different walkers in three London areas and Gävle

|  |  | Barnsbury | Clerkenwell | S. Kensington | Mean | Gävle |
|---|---|---|---|---|---|---|
| Random walkers | WPR | 0.87 | 0.92 | 0.91 | **0.90** | **0.89** |
|  | PR  | 0.86 | 0.87 | 0.84 | **0.86** | **0.75** |
|  | Cnt | 0.86 | 0.87 | 0.84 | **0.86** | **0.75** |
|  | Ctr | 0.54 | 0.48 | 0.44 | 0.48 | **0.41** |
|  | Btw | 0.67 | 0.64 | 0.57 | 0.63 | **0.56** |
|  | Ltg | 0.82 | 0.83 | 0.80 | **0.82** | **0.36** |
|  | Gtg | 0.23 | 0.32 | 0.27 | 0.27 | 0.26 |
| Purposive walkers (I) | WPR | 0.63 | 0.63 | 0.57 | **0.61** | **0.67** |
|  | PR  | 0.59 | 0.58 | 0.53 | **0.56** | **0.57** |
|  | Cnt | 0.59 | 0.58 | 0.53 | **0.56** | **0.57** |
|  | Ctr | 0.42 | 0.42 | 0.38 | 0.41 | **0.42** |
|  | Btw | 0.81 | 0.77 | 0.75 | **0.77** | **0.57** |
|  | Ltg | 0.54 | 0.51 | 0.46 | **0.50** | **0.16** |
|  | Gtg | 0.20 | 0.19 | 0.16 | 0.19 | 0.13 |
| Purposive walkers (II) | WPR | 0.73 | 0.73 | 0.68 | **0.71** | **0.75** |
|  | PR  | 0.69 | 0.69 | 0.64 | **0.68** | **0.60** |
|  | Cnt | 0.69 | 0.69 | 0.64 | **0.68** | **0.60** |
|  | Ctr | 0.47 | 0.46 | 0.41 | 0.45 | **0.33** |
|  | Btw | 0.80 | 0.77 | 0.74 | **0.77** | **0.53** |
|  | Ltg | 0.67 | 0.63 | 0.59 | **0.63** | **0.33** |
|  | Gtg | 0.20 | 0.20 | 0.17 | 0.19 | 0.24 |

Table 5: R square values between the metrics and the simulated flow (using gate counts) for different walkers in three London areas and Gävle

|  |  | Barnsbury | Clerkenwell | S.Kensington | Mean | Gävle |
|---|---|---|---|---|---|---|
| Random walkers | WPR | 0.90 | 0.92 | 0.90 | **0.91** | **0.71** |
|  | PR  | 0.91 | 0.89 | 0.87 | **0.89** | **0.68** |
|  | Cnt | 0.91 | 0.89 | 0.87 | **0.89** | **0.68** |
|  | Ctr | 0.58 | 0.47 | 0.45 | 0.50 | **0.42** |
|  | Btw | 0.64 | 0.59 | 0.51 | 0.58 | **0.47** |
|  | Ltg | 0.83 | 0.83 | 0.80 | **0.82** | **0.34** |
|  | Gtg | 0.24 | 0.32 | 0.24 | 0.26 | 0.16 |
| Purposive walkers (I) | WPR | 0.67 | 0.67 | 0.60 | **0.65** | **0.52** |
|  | PR  | 0.68 | 0.66 | 0.63 | **0.65** | **0.58** |
|  | Cnt | 0.68 | 0.66 | 0.63 | **0.65** | **0.58** |
|  | Ctr | 0.51 | 0.49 | 0.49 | 0.50 | **0.48** |
|  | Btw | 0.83 | 0.80 | 0.79 | **0.81** | **0.58** |
|  | Ltg | 0.61 | 0.57 | 0.51 | **0.56** | **0.28** |
|  | Gtg | 0.22 | 0.22 | 0.16 | 0.20 | 0.11 |
| Purposive walkers (II) | WPR | 0.78 | 0.77 | 0.73 | **0.76** | **0.56** |
|  | PR  | 0.80 | 0.78 | 0.77 | **0.78** | **0.67** |
|  | Cnt | 0.80 | 0.78 | 0.77 | **0.78** | **0.67** |
|  | Ctr | 0.59 | 0.56 | 0.55 | 0.57 | **0.53** |
|  | Btw | 0.81 | 0.79 | 0.76 | **0.79** | **0.62** |
|  | Ltg | 0.72 | 0.67 | 0.64 | **0.68** | **0.38** |
|  | Gtg | 0.21 | 0.22 | 0.17 | 0.20 | 0.15 |

Why can integration or closeness centrality not capture human movement rates? Examination of these metrics and the aggregate flow found that all of them demonstrate a power law or power law-like distribution (Jiang, Zhao and Yin 2008). However, this is an exception for the closeness centrality metric. In other words, closeness centrality is not power law distributed, but rather normally distributed. This is the very reason why in the above metric-flow correlation we took quartic root in order to transform the metrics from a power law distribution to a normal distribution. Given the above argument, why can Ltg capture aggregate flow? Ltg by definition is very similar to Cnt, and it is just one step further in computing its values. In other words, Cnt considers only friends, while Ltg considers both friends, and friends of the friends. It is in this sense Ltg sometimes appears to be one of the best indicators, in particular in the London cases.

To this point, we have seen that both observed flow and simulated flow are highly correlated to the metrics. In other words, street structure has a great impact on both observed and simulated movement. We can hypothesize that both observed flow and simulated flow should be highly correlated as well, for they appear to have the same cause – the underlying street structure. For the London cases, there are observed flow for both pedestrians and vehicles, so it is possible to verify the hypothesis. Table 6 indeed illustrates the substantial correlation between the observed and the simulated. Among the three kinds of moving agents, random walkers appear to be the best, followed by purposive walkers (II) and then purposive walkers (I).

Table 6: Correlation between the observed and simulated flows

|  | Barnsbury | | Clerkenwell | | S.Kensington | | Knightsbridge | | Mean | |
| --- | --- | --- | --- | --- | --- | --- | --- | --- | --- | --- |
|  | Ped | Veh | Ped | Veh | Ped | Veh | Ped | Veh | Ped | Veh |
| Random walkers | 0.89 | 0.82 | 0.72 | 0.80 | 0.76 | 0.87 | 0.66 | 0.73 | 0.76 | 0.81 |
| Purposive walkers (I) | 0.82 | 0.74 | 0.49 | 0.76 | 0.55 | 0.73 | 0.65 | 0.65 | 0.63 | 0.72 |
| Purposive walkers (II) | 0.86 | 0.78 | 0.61 | 0.82 | 0.60 | 0.80 | 0.69 | 0.68 | 0.69 | 0.77 |

From the above results, we can assert two important points: (1) human moving behavior has little effect on the aggregate flow, and thus the movement patterns (at a collective level) formed by human beings and random walkers are essentially the same, (2) closeness centrality is not a good indicator for aggregate flow. In space syntax research, seeking correlation between integration and traffic flow or any other phenomenon has been a regular exercise. Once the link is found, we say that morphological metrics "explain" the other. Inquiry often stops here, without exploring in detail the basic causal mechanism into view. Herewith through agent-based simulation, we made some deep insights toward the understanding of human movement patterns.

### 5. Conclusion
We relied on different moving agents to explore the mechanisms underlying the emergence of human movement patterns in street networks. The moving agents differ fundamentally in their moving behavior: one random and two purposive. However, the movement patterns formed by the different agents remain similar. This implies that the movement patterns formed by human beings (purposive walkers) and random walkers are the same. It demonstrates that higher cognitive abilities are not required in the formation of movement patterns at a collective level. This is opposite to Hillier's view of cognitive effects. The movement patterns in street networks are self-organized through the interaction between moving agents and the underlying street networks, and have little to do with agents' cognition. We further examined the disparities among the metrics and found that closeness centrality is not a good indicator for predicting traffic flow. Even the local integration is far from the best indictor. Instead we suggest some alternatives: Google's PageRank, its modified version - weighted PageRank, betweenness and connectivity centralities.

Beyond these findings, our work has added an interesting example to the study of emergence of collective behavior (Goldstone and Gureckis 2009), which often has some quite simple origins (Ball 2004, Buchanan 2007). In this respect, agent-based simulation is a powerful tool, focusing on the simple interaction between the agents and the streets from the bottom up. Many current studies have been devoted to verifying the powerfulness of the method per se, but few studies concentrate on what useful patterns or knowledge we can discover from using agent-based simulation. This paper provides a compelling example of how the kind of agent-based simulation can be used to uncover the mechanisms of human movement patterns. We share this view with Epstein (2008) that this model function, together with many others, has been overlooked in studying complex social and spatial phenomena. Our model is simple enough, yet captures the essence of complicated movement patterns. Our future work will extend the approach to exploring other complex collective behaviors in geographic space.

**Acknowledgement**
This paper was initially submitted to the Annals of the Association of American Geographers, and was rejected after one month or so by an editor without sending out for reviewing process. While admitting that it is highly innovative and interesting, the editor considered the paper to be geography-unrelated. Thanks for the editor's decision that leads to this publication. We also would like to thank the three reviewers of IJGIS for their constructive comments that better shaped the paper.

**References:**
Ball P. (2004), *Critical Mass: how one thing leads to another*, William Heinemann: London.
Buchanan M. (2007), *The Social Atom: why the rich get richer, cheaters get caught, and your neighbor usually looks like you*, Bloomsbury: New York.
Epstein J. M. (2008), Why model? *Journal of Artificial Societies and Social Simulation*, 11.4, 12
Figueiredo L. and Amorim L. (2005), Continuity lines in the axial system, in van Nes, A. (ed.), *Proceedings of the 5th International Symposium on Space Syntax*, TU Delft, Delft, Netherlands, 163–174.
Freeman, L. C. (1979). Centrality in social networks: Conceptual clarification, *Social Networks*, 1(3), 215-239.
Goldstone R. L. and Gureckis T. M. (2009), Collective behavior, *Topics in Cognitive Science*, 3.1, 412-438.
Gonzalez M., Hidalgo C. A., and Barabási A.-L. (2008), Understanding individual human mobility patterns, *Nature*, 453, 779 – 782.
Hillier B. (1999), The hidden geometry of deformed grids: or, why space syntax works, when it looks as though it shouldn't, *Environment and Planning B: Planning and Design*, 26, 169-191.
Hillier B. and Hanson J. (1984), *The Social Logic of Space*, Cambridge University Press: Cambridge.
Hillier, B. (1996), *Space Is the Machine: a configurational theory of architecture*, Cambridge University Press: Cambridge.
Hillier, B. and Iida, S. (2005), Network and psychological effects in urban movement. In: Cohn, A.G. and Mark, D.M. (eds.), *Proceedings of the International Conference on Spatial Information Theory*, COSIT 2005, Ellicottsville, N.Y., U.S.A., September 14-18, 2005, Springer-Verlag: Berlin, 475-490.
Hillier, B., Penn A., Hanson J., Grajewski T. and Xu J. (1993), Natural movement: configuration and attraction in urban pedestrian movement, *Environment and Planning B: Planning and Design*, 20, 29-66.
Jiang B. (2007), A topological pattern of urban street networks: universality and peculiarity, *Physica A: Statistical Mechanics and its Applications*, 384, 647 - 655.
Jiang B. (2009), Ranking spaces for predicting human movement in an urban environment, *International Journal of Geographical Information Science*, 23.7, 823–837, Preprint, arxiv.org/abs/physics/0612011.
Jiang B. and Claramunt C. (2004), Topological analysis of urban street networks, *Environment and Planning B: Planning and Design*, 31, 151- 162.
Jiang B. and Liu C. (2009), Street-based topological representations and analyses for predicting traffic flow in GIS, *International Journal of Geographical Information Science*, 23.9, 1119–1137. Preprint, arxiv.org/abs/0709.198.
Jiang B., Yin J., and Zhao S. (2009), Characterizing human mobility patterns in a large street network, *Physical Review E*, 80, 021136，Preprint, arXiv:0809.5001.
Jiang B., Zhao S., and Yin J. (2008), Self-organized natural roads for predicting traffic flow: a sensitivity study, *Journal of Statistical Mechanics: Theory and Experiment*, July, P07008, Preprint, arxiv.org/abs/0804.1630.
Langville A. N. and Meyer C. D. (2006), *Google's PageRank and Beyond: the science of search engine rankings*, Princeton University Press: Princeton, N.J.
Penn, A. and Dalton, N. (1994), The architecture of society: stochastic simulation of urban movement, in N. Gilbert and J. Doran (eds), *Simulating Societies: The Computer Simulation of Social Phenomena*, UCL Press, London, pp. 85–125.
Page L., and Brin S. (1998), The anatomy of a large-scale hypertextual Web search engine, *Proceedings of the seventh international conference on World Wide Web* 7, 107-117.
Penn A., Hillier B., Banister D. and Xu J. (1998), Configurational modeling of urban movement networks, *Environment and Planning B: Planning and Design*, 25, 59-84.
Thomson R. C. (2003), Bending the axial line: smoothly continuous road centre-line segments as a basis for road network analysis, in Hanson, J. (ed.), *Proceedings of the Fourth Space Syntax International Symposium*, University College London, London.
Turner A. (2007a), To move through space: lines of vision and movement, in Kubat, A. S. (ed.), *Proceedings of the 6th International Symposium on Space Syntax*, Istanbul Teknik Üniversitesi, Istanbul.
Turner A. (2007b), From axial to road-centre lines: a new representation for space syntax and a new model of route choice for transport network analysis, *Environment and Planning B: Planning and Design*, 34, 539-555.
Turner A. (2009), The role of angularity in route choice: an analysis of motorcycle courier GPS traces. In: Stewart Hornsby K. and Claramunt C. and Denis M. and Ligozat G. (eds.), *Spatial Information Theory*, Springer Verlag, Berlin/ Heidelberg, Germany, 489-504.
Wilensky U. (1999), *NetLogo*, http://ccl.northwestern.edu/netlogo/. Center for Connected Learning and Computer-Based Modeling, Northwestern University. Evanston, IL.
Xing W. and Ghorbani A. (2004), Weighted PageRank algorithm, in: Ali A. Ghorbani (editor), *Second Annual Conference on Communication Networks and Services Research CNSR'04*, Fredericton, N.B. Canada, May

19-21, 2004, 305-314.

**Appendix A: A naïve introduction to the seven metrics for characterizing street structure**

The seven metrics, including weighted PageRank (WPR), PageRank (PR), connectivity (Cnt), control (Ctr), betweenness (Btw), local integration (Ltg), and global integration (Gtg), have been well studied in the literature to characterize street structure. Many of them were developed initially in social networks as the concept of centrality. To make the paper self-contained, we make a naïve introduction to the metrics, and leave more details and mathematics to the reader to refer to in the relevant literature (e.g., Freeman 1979, Jiang 2009). Let us use the kite-shaped graph (Figure A) to explain the ideas and motivations behind the metrics. If we take node j for example, its Cnt = 1. This is because it is connected to node i only. The control value of a node is the sum of the reciprocal of its neighbors' connectivity. For node i, its $Ctr = \frac{1}{1} + \frac{1}{3}$, since its two neighbors have a connectivity of 1, and 3 respectively. Connectivity considers only directly connected nodes. What if we consider both directly and indirectly connected nodes? This is the basic notion of closeness centrality (or global integration), i.e, the extent to which a node is connected to all the other nodes in a graph. Taking node j for example, its Gtg = $1\times1+1\times2+2\times3+5\times4$, for it has 1 directly connected node (with step 1), 1 connected node with step 2, 2 connected nodes with step 3, and 5 connected nodes with step 4. Instead of considering all directly and indirectly connected nodes, we are sometimes only interested in those nodes within a few steps. This is the idea of local integration. For node j, its Ltg =$1\times1+1\times2$, considering those nodes within two steps. Both connectivity and local integration are local metrics, while global integration is a global metric. A third perspective of centrality is to assess how a node is between two parts, a sort of bridging role. For example, node h has the highest betweenness value followed by f, g and i. Without the nodes of the highest betweenness, the kite graph would very likely be broken into two pieces.

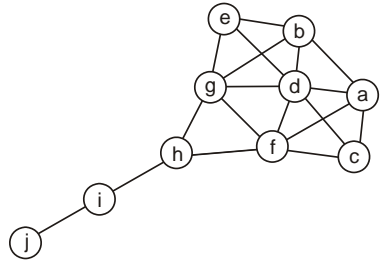

Figure A: A kite-shaped graph

PageRank is justified as a random surfer who randomly jumps from web page to web page by arbitrarily clicking or following hotlinks (Page and Brin 1998). Eventually the visiting frequency of an individual web page can be characterized by the PageRank score. Assuming that the kite graph is a web graph in which nodes represent individual web pages, while links are hotlinks, if the random surfer is currently at node h, it has three choices from which it randomly can jump to. In other words, the surfer gives the same priority to all the hot linked pages. It is important to note that a web graph is a directed rather than undirected graph unlike those we have adopted so far. A web graph also includes dangling nodes which have no hotlinks, such as images. Given these particularities, the definition of PageRank is rather more complicated than what we have presented here; For more details about the definition and mathematics, we refer the reader to Langville and Meyer (2006) or Jiang (2009) in a brief form. In fact, for an undirected and fully connected graph without dangling nodes, PageRank scores ranking are not very different from that of connectivity. This is why we have seen that connectivity stays with PageRank at almost identical levels in the above experiments. Weighted PageRank differentiates from the standard PageRank in only one aspect, that is, in choosing the next page, WPR gives a higher priority to those highly connected pages (Xing and Ghorbani 2004). For example, if one is at node h, a priority is given to nodes f and g rather than i for the next jumping; In contrast, in standard PageRank, no such priority bias is introduced among the connected nodes f, g, and i.

In summary, we could group the seven metrics into four categories: (1) WPR, (2) PR, Cnt, and Ctr, which are all connectivity related, (3) Btw, and (4) Ltg and Gtg. Each category shows a great similarity in terms of definition and how they perform in capturing traffic flow. It should be noted that local integration sometimes can be put into group two, as it is by definition closer to connectivity.

**Appendix B: Algorithms for simulating network constrained movement**
(Refer to the site http://fromto.hig.se/~bjg/movingbehavior/ for a demo showing the random and purposive moving behaviours)

```
Algorithm I: Procedure for random agents
----------------------------------------------------------------------
Input: A street network, number of agents, and simulation time
Output: aggregate flow (footprints and gatecounts)

Function Random_Movement (streetNetwork, aNumber, sTime)
    Construct the connectivity graph from streetNetwork;
    Generate agents(aNumber), locate the agents on the streets randomly, and set their next streets;
    While ticks <= sTime
        For each agent in agents
            If agent reaches the next street
                Select the next street from its connected streets randomly with a priority on
                the most connected street based on the connectivity graph;
            Else
                Move the agent along the current street towards the next street;
                Count aggregate flow by adding 1 to the footprints on the current road;
                Count aggregate flow by adding 1 to a gate if the agent passed through one gate;
        ticks = ticks + 1;

Algorithm II: Procedure for two types of purposive agents
----------------------------------------------------------------------
Input: A street network, number of agents, purpose choice and simulation time
Output: aggregate flow (footprints and gatecounts)

Function Purposive_Movement (streetNetwork, aNumber, sTime, pChoice)
    Construct the connectivity graph from streetNetwork;
    Generate agents(aNumber), and locate the agents on the streets randomly;
    For each agent in agents
        Purposive_walker_behavor (streetNetwork, pChoice);
        Compute the shortest distance path between the current location and destination;
    While ticks <= sTime
        For each agent in agents
            If agent reaches the destination
                Purposive_walker_behavor (streetNetwork, pChoice);
                Compute the shortest distance path between the current location and destination;
            Else
                Move the agent along the shortest path towards the destination;
                Count aggregate flow by adding 1 to the footprints on the current road;
                Count aggregate flow by adding 1 to a gate if the agent passed through one gate;
        ticks = ticks + 1;

//Different purposive movement behaviors
//Return a destination
Function Purposive_walker_behavior (streetNetwork, pChoice)
    If pChoice = I
        Select a destination street from streetNetwork;
    Else if pChoice = II
        nearStreets = Streets within two steps of topological distance from the current street
        based on the connectivity graph;
        farStreets = streetNetwork - nearStreets;
        Select a destination street from nearStreets with probability 80% and farStreets with
        probability 20%;
    Take any location on the destination street as the destination to target;
```